\documentstyle[12pt,times,epsfig]{article}

\makeatletter                    
\@addtoreset{equation}{section}  
\makeatother                     

\newskip\humongous \humongous=0pt plus 1000pt minus 1000pt

\newif\ifdtup

\begin{document}

\title{Path Dependent Option Pricing: the path integral partial averaging method}
\author{Andrew Matacz$^1$\thanks{Email: andrew.matacz@science-finance.fr}\\
{\small School of Mathematics and Statistics}\\
{\small University of Sydney, 2006} \\
{\small Australia}\\
{\small $^1$Current address: Science \& Finance} \\
{\small 109-111 rue Victor Hugo}\\
{\small 92532 Levallois, France}}
\maketitle

\begin{abstract}
In this paper I develop a new computational method for 
pricing path depend\-ent options. Using the path integral 
representation of the option price, I show that in general it is possible 
to perform analytically a partial averaging over the underlying 
risk-neutral diffusion process. This result greatly eases the computational 
burden placed on the subsequent numerical evaluation. 
For short-medium term options it leads to a general approximation 
formula that only requires the evaluation of a one dimensional 
integral. I illustrate the application of the method to Asian options 
and occupation time derivatives.

\end{abstract}

\newpage
\section{Introduction}

Financial derivatives (eg. options and futures) derive their value
from an underlying traded financial security, whose price is modeled
by some stochastic process. In their most general form, the option payoff 
is path dependent since it depends on the entire future path traversed 
by the underlying security. Path dependent options are defined using 
either discrete or continuous price sampling. For continuous sampling 
closed from solutions are often available, but in practice most traded path 
dependent options are discretely sampled. It is known that the application 
of these closed form solutions leads to substantial pricing errors for 
discretely sampled options \cite{cha,kat,lev}. This feature has 
necessitated the development of practical and efficient computational 
methods for the evaluation of path dependent options \cite{cle}.
Most research has focussed on either partial differential
equation, Monte Carlo or tree based methods. In contrast to these 
approaches, in this paper I will develop an alternative based 
on the {\it path integral} formulation of the pricing problem. 

For many years, theoretical physicists have been developing and
applying the path integral method for calculating expectations
similar to those now being encountered in the evaluation of financial
derivatives (via the risk-neutral valuation formula). 
A path integral is an infinite dimensional Riemannian
integral as the integral is performed over a set of functions or
paths. The path integral method is unique in that it gives a global
formulation of the problem in question. 
This global description provides a
powerful tool for deriving analytical approximation and
numerical solution schemes that are difficult or impossible to
formulate in other ways. Formally, the path integral method is easily
extended to multi-dimensional problems. Much of the driving force
behind the development of path integral numerical methods in physics
has been that, compared to other methods, they show only a slow
increase in computational complexity as the dimensionality of the
problem increases. Path integral methods were first introduced by
Feynman in 1942 as an alternative formulation of quantum physics
\cite{feynman,grosche}. They have found wide application in the
evaluation of both the real time dynamics and equilibrium statistical
mechanics of quantum many-body systems \cite{makri,drozdov1,negele}.
They are also a popular and natural tool for the analysis of diffusion
processes \cite{strat,drozdov2,drozdov3}, including non-Markovian
systems where there is a lack of practical alternatives \cite{hugh}.

The application of path integral methods to financial derivatives was
pioneered by Dash who developed a path integral
framework for pricing bonds and options within one factor term
structure models \cite{dash1,dash2,dash3}. More recently, Linetsky
\cite{linetsky1} was the first to show how path dependent options
could be formally priced in a path integral framework. He considered several
examples of one and two factor path dependent options. Baaquie
\cite{baaquie1} has shown how path integral methods can be used to
price vanilla options with stochastic volatility. The
same author has also recast the popular Heath-Jarrow-Morton model of forward 
interest rates as a problem in path integration \cite{baaquie2}. 
Similar to Dash, Otto has more recently shown how to use path integration to price bonds and
bond options under general short rate models \cite{otto}. Bennati {\it et-al} 
\cite{rosa-clot1} have focussed on a multi-dimensional path integral 
formalism for solving general financial problems based on systems of 
stochastic equations.
All these preceding authors focussed on general formalism and exactly 
solvable models. They have shown that path integrals constitute a natural framework for
describing the evaluation of general multi-factor derivatives.
Although path integral methods offer an attractive way of obtaining
exact solutions, they cannot find exact solutions not available using
more standard methods. The greatest promise of path integral methods
will be in the development of new numerical and approximation methods
for addressing pricing problems where exact
solutions are impossible.

Path integral numerical methods involve the evaluation of a
multi-dimensional integral, either by deterministic or Monte Carlo 
methods. A review of deterministic and
Monte Carlo methods, developed for physics applications, can be found in
Drozdov \cite{drozdov2} and Makri \cite{makri} respectively. An
independent and much more recent development has been the application
of path integral numerical methods to financial derivatives. This was
pioneered by Eydeland \cite{Eyd} who, using fast Fourier transform
methods, has derived a deterministic path integral algorithm for
calculating the generating function of a random variable defined as
the time integral of a general diffusion process. He pointed to a
number of potential financial applications of this algorithm.
Chiarella and El-Hassan have applied the method of Eydeland to the
pricing of American bond options in the Heath-Jarrow-Morton framework
\cite{chiarella1}. More recently, Chiarella {\it et-al} have devised a
deterministic path integral algorithm based on Fourier-Hermite series
expansions and applied it to the pricing of American options
and point barrier options \cite{chiarella2,chiarella3}. 
They report computational times that are a significant improvement 
over the standard binomial and finite difference methods. 
In contrast to these deterministic methods, Makivic \cite{makivic} has 
initiated
research into the path integral Monte Carlo evaluation of 
financial derivatives. He broadly
outlines a computational approach based on the standard Metropolis
algorithm. Rosa-clot and Taddei \cite{rosa-clot2} have discussed both a 
deterministic method and the Monte Carlo approach and applied them to several 
examples. Some of these previous papers point out that the path integral 
method is superior to the the traditionally used 
lattice methods, because the underlying asset price is left continuous rather
than being discretized. This has several important advantages. The
option price is obtained more accurately since all possible price
paths are included in the simulation. Option Greeks are obtained more
reliably since the method avoids the need for numerical
differentiation. A further advantage of path integral methods is that
they can be easily and efficiently extended to evaluate multi-factor
financial derivatives.

The application of path integral methods here is fundamentally 
different from all the previous cited works. We use the path integral 
representation of the option price to show that rather generally, it is possible 
to perform analytically a {\it partial averaging} over the underlying 
risk-neutral diffusion process. This key result will greatly 
reduce the computational burden placed on the subsequent numerical evaluation. 
The application of this method is inspired by a technique 
first developed for computation problems in chemical physics 
\cite{coalson1,coalson}. Conceptually, the partial averaging corresponds 
to averaging over the high frequency fluctuations of the 
risk-neutral diffusion process. For short-medium term options it leads to 
a general approximation formula that only requires the evaluation of a 
one dimensional numerical integral. Longer term options can be evaluated 
deterministically or most generally by standard Monte Carlo methods. 
In this case, the partial averaging method will greatly reduce the required 
dimension of the Monte Carlo simulation.

The outline of this paper is as follows. In section 2 we develop 
a formal path integral representation for the evaluation of rather 
general path dependent options. In section 3 we show how the previous
path integral can be numerically evaluated after performing 
analytically a partial averaging over the 
underlying risk-neutral diffusion process. In section 4 some examples 
are presented. For clarity of presentation and completeness, the path integral 
methods used in this paper are developed from first principles in the 
three appendices.

\section{Path dependent option theory}
In this section we will present a formal path integral 
framework general enough to price a wide range of path dependent options.
We will assume the standard geometric Brownian motion ({\sc gbm})
model of the asset price process. More general diffusion processes 
present no special difficulties. 
In this case the risk-neutral price process is given by the Ito stochastic 
differential equation
\begin{equation}
dS_{t}= (r-q)S_tdt+\sigma S_{t}dW_{t},
\end{equation}
where $r$ is the interest rate and $q$ is the continuous dividend yield.
Using Ito's lemma we can show
\begin{equation}
dx_{t}=\mu dt+\sigma dW_{t},
\end{equation}
where
\begin{equation}
x_t=\ln S_t,\;\;\;\mu=r-q-\frac{1}{2}\sigma^2.
\end{equation}

Consider a path dependent option with price $C_F$ at expiry $u$ given by
\begin{equation}
C_F(S_u,{\cal I},u)=F(S_u,{\cal I})=F(e^{x_u},{\cal I}),
\end{equation}
where $F$ is the option payoff function which depends on some
path dependent random variable ${\cal I}$. In this paper we 
will assume that ${\cal I}$ can be written as
\begin{equation}
{\cal I}=\int_t^u ds\,w(s)f(x_s,s),
\end{equation}
which is a time integral over an arbitrary function of the risk-neutral 
diffusion process (2.2). For continuous sampling $w(s)=1$, but for discrete sampling
(which is more realistic in practice)
\begin{equation}
w(s)=\sum_{i} w_i\delta(s-s_i),
\end{equation}
where $w_i$ are the sampling weights and $s_i$ are the sampling times.
The above definition of ${\cal I}$ was used before by 
Wilmott {\it et-al} \cite{wil} whose focus was on partial differential 
equation methods. It was shown to be general enough to include Asian, 
barrier and lookback options which are 3 qualitatively different path 
dependent options. 
We will present examples in section 4.

In a risk-neutral framework, the option price at inception time $t$ 
is given by
\begin{equation}
C_F(S_t,t)=e^{-rT}E_{x_t}[F(e^{x_u},{\cal I})],
\end{equation}
where $T=u-t$ and the expectation is with respect to the transformed
risk-neutral price process (2.2) conditioned on the initial value $x_t$.
The price at inception can then be expressed as
\begin{equation}
C_F(S_t,t)=e^{-rT}\int^{\infty}_{-\infty}dx_u\int^{\infty}_{-\infty}
d{\cal I}\,P(x_u,{\cal I}|x_t)F(e^{x_u},{\cal I}),
\end{equation}
where $P(x_u,{\cal I}|x_t)$ is the joint probability density function 
({\sc pdf}) of $x_u$ and the path dependent
random variable ${\cal I}$. In appendix 
A we show for a general diffusion process how the joint {\sc pdf} can be formally 
computed as a path integral. For the 
special case of {\sc gbm},
we show in appendix A that the joint {\sc pdf} is given by
\begin{equation}
P(x_u,{\cal I}|x_t)=\frac{1}{2\pi}\exp\left[\frac{\mu x_u}{\sigma^2}
-\frac{\mu x_t}{\sigma^2}-\frac{\mu^2 T}{2\sigma^2}\right]
\int_{-\infty}^{\infty}dk e^{-ik{\cal I}}K(x_u,x_t;T),
\end{equation}
where $\mu$ is the constant defined in (2.3) and
$K$, which we refer to as the propagator is defined by 
\begin{equation}
K(x_{u},x_{t};T)=\int_{x_{t}}^{x_{u}}{\cal D}x_s\exp
\left[ -\frac{1}{2\sigma ^{2}}\int_{t}^{u}ds\,\,\left( \dot{x}^{2}_s
+V(x_s,s)\right) \right].
\end{equation}
In (2.10), the integration measure ${\cal D}x_s$ denotes a {\em path integral} which is 
defined precisely in appendix A. It describes an infinite dimensional integral 
over all paths connecting $x_u$ at the expiry time and $x_t$ at the initial 
time. 
We refer to the function $V$ in (2.10) as the potential function.
For the {\sc gbm} model it has the form
\begin{equation}
V(x_s,s)=-2ik\sigma ^{2}\,w(s)\,f(x_s,s).
\end{equation}
We see that in this case the potential is imaginary with a functional 
form determined by the path dependent random variable (2.5). We show in 
appendix A that for more general risk-neutral diffusion processes the 
potential function is complex.

The propagator (2.10), up to a boundary term, is the characteristic
function of the joint {\sc pdf}. In physics, 
the path integral (2.10) is equivalent to the path integral representation
for the equilibrium quantum statistical density matrix of a particle in a
complex potential $V(x)$ \cite{feynman}.
In this case temperature replaces the role of
time in (2.10).
Under an imaginary time transformation, (2.10) becomes equivalent to that
which gives the quantum mechanical propagator of a one dimensional
quantum particle in a complex potential $V(x)$.
These identifications with standard problems in theoretical physics are
of great value because we can then use the methods and results
of theoretical physics for performing these path integrals.
Tables of known exact path integrals of the form (2.10), for various potential
functions, are listed by Grosche \cite{grosche}. Such tables along with
the formulation provided here provide an easy way to obtain exact solutions
to path dependent option prices.

\subsection{Seasoned path dependent options}
In this paper we consider the option at its inception time.
More generally, the option price at time $t'$ ($t<t'<u$) for a
seasoned path dependent option is given by
\begin{equation}
C_F(S_{t'},{\cal I}_{t}^{t'},t')=e^{-r(u-t')}E_{x_{t'}}
[F(e^{x_u},{\cal I}_t^{t'}+{\cal I}_{t'}^u)],
\end{equation}
where we use the more detailed notation
\begin{equation}
{\cal I}_{t}^u=\int_t^u ds\,w(s)f(x_s,s).
\end{equation}
We then find
\begin{equation}
C_F(S_{t'},{\cal I}_t^{t'},t')=e^{-r(u-t')}\int^{\infty}_{-\infty}
dx_u\int^{\infty}_{-\infty}
d{\cal I}_{t'}^u\,P(x_u,{\cal I}_{t'}^u|x_{t'})
F(e^{x_u},{\cal I}_{t}^{t'}+{\cal I}_{t'}^u).
\end{equation}
We see that this case only affects the payoff function in a simple way and 
the joint {\sc pdf} we need to find is the same problem as before. Therefore all 
the final results can be trivially extended to seasoned options.

\section{Partial Averaging}
In the previous section we showed that for the {\sc gbm} model, 
equations (2.5) and (2.9-11) define a formal path integral representation 
for the joint {\sc pdf}. The option 
price is then obtained from this joint {\sc pdf} via (2.8).
In this section we use the previous path integral formulation
to show that it is possible to perform analytically a partial 
averaging \cite{coalson1} over the underlying risk-neutral 
diffusion process. The option price can then be more efficiently 
evaluated by numerical methods.

First we must discretize in time (2.2), by defining the discrete time 
$s_{n}=n\varepsilon +t$,
where $n=0,1,...,N$ and $\varepsilon =T/N$ with $T=u-t$.
The propagator (2.10) can then be decomposed as
\begin{equation}
K(x_{u},x_{t};T)=\int^{\infty}_{-\infty}
dx_{N-1}...dx_1\prod_{n=1}^{N}K(x_n,x_{n-1};\varepsilon)
\end{equation}
where $x_N\equiv x_u,\;x_0\equiv x_t$ and a general form for the 
short-time propagator is, as shown in appendix C,
\begin{equation}
K(x_n,x_{n-1};\varepsilon)=\left(\frac{1}{2\pi\sigma^2\varepsilon}
\right)^{1/2}\exp\left[-\frac{(x_n-x_{n-1})^2}{2\sigma^2\varepsilon}
-\gamma(x_n,x_{n-1};\varepsilon)\right],
\end{equation}
where $\gamma$ will be defined below.
Substituting (3.2) into (3.1) we find that the propagator becomes
\begin{equation}
K(x_{u},x_{t};T)= \left(\frac{1}{2\pi\sigma^2\varepsilon}\right)^
{\frac{N}{2}}\int^{\infty}_{-\infty}
dx_{N-1}...dx_1\exp\left[-\sum_{n=1}^N\frac{(x_n-x_{n-1})^2}
{2\sigma^2\varepsilon}-\sum_{n=1}^N\gamma(x_n,x_{n-1};\varepsilon)
\right].
\end{equation}
As $\varepsilon\rightarrow 0$, its possible to show that
\begin{equation}
\gamma(x_n,x_{n-1};\varepsilon)=
\varepsilon\frac{(V(x_n,s_n)+V(x_{n-1},s_{n-1}))}{4\sigma^2}
\end{equation}
yields the correct short-time propagator when substituted into (3.2).

We will refer to equation (3.2), with (3.4), as the primitive short-time 
propagator. It is the standard short-time propagator used 
in the numerical evaluation of path integrals.
A key feature of this propagator is that it is {\it not} 
correct to first order in $\varepsilon$. It is in fact only valid as
$\varepsilon\rightarrow 0$. Clearly, the dimension of the integral in 
(3.1) could be made much smaller by searching for short-time propagators 
accurate over larger time-steps. This observation has motivated the 
search for improved short-time propagators for use in path integral 
calculations for physics applications \cite{makri,drozdov4,coalson}.
In appendix C, we show that its possible in general to write
\begin{equation}
\gamma(x_n,x_{n-1};\varepsilon)=
-\sum_{m=1}^{\infty}\frac{1}{m!}
\left(-\frac{\varepsilon}{2\sigma^2}\right)^m
C_m(x_n,x_{n-1};\varepsilon),
\end{equation}
where $C_m$ describes a cumulant structure with each cumulant of order
$(\sigma^2\varepsilon)^{m-1}$. The key result is, we can obtain a 
short-time propagator formally correct to {\it second} order in $\varepsilon$, 
by truncating all cumulants beyond the first. This truncation 
corresponds to retaining an averaging over only the high frequency fluctuations 
of the risk-neutral diffusion process; i.e. a partial averaging. 
The improved short-time propagator 
will lead to a much more efficient numerical evaluation compared to that 
obtained by using the primitive short-time propagator.
In appendix C, we calculate the first 2 cumulants 
exactly for a general potential and derive an expansion of the propagator 
to third order in $T$ (a result only valid for smooth potentials).

Using the results from appendix C, with the {\sc gbm} potential (2.11), 
we find that
\begin{equation}
\gamma(x_n,x_{n-1};\varepsilon)\simeq -i\varepsilon k
\alpha(x_n,x_{n-1};\varepsilon)+\frac{\varepsilon^2 k^2}{2}
\beta(x_n,x_{n-1};\varepsilon)+o(\sigma^4\varepsilon^5),
\end{equation}
where
\begin{equation}
\alpha(x_n,x_{n-1};\varepsilon)=
\int_0^1d\tau\,w(\tau)\int^{\infty}_{-\infty}dp_{\tau}\,P(p_{\tau})
f(\bar{x}_\tau+p_{\tau},\tau),
\end{equation}
\begin{equation}
P(p_{\tau})=\frac{1}{\sqrt{2\pi\nu_\tau^2}}
\exp(-p_{\tau}^2/2\nu_\tau^2)
\end{equation}
and
\begin{equation}
\nu_\tau^2=\sigma^2 \varepsilon(1-\tau)\tau,\;\;
\bar{x}_{\tau}=\tau(x_n-x_{n-1})+x_{n-1},\;\;
\tau=(s-s_{n-1})/\varepsilon.
\end{equation}
Equation (3.7) is a key equation as it contains the partial averaging.
The origin of $\alpha$ is the first cumulant in the expansion 
(3.5) and its evaluation will lead to a short-time propagator 
correct to second order in $\varepsilon$. 
Expanding $\alpha$ to order $\varepsilon$ is consistent with the 
truncation of the higher cumulants in (3.5). 
The origin of $\beta$ is the second cumulant which is formally
calculated in appendix C. All we need to know here is that $\beta$ is
of order $\sigma^2\varepsilon$, since it will be set to zero at the end. 
We keep it to determine the order of the first correction term due to 
the truncation of all higher cumulants.
After combining (3.6),(3.3) and (2.9) and performing the 
integration over $k$, we find that the joint {\sc pdf} becomes
\begin{eqnarray}
P(x_u,{\cal I}|x_t)&\simeq&
\int^{\infty}_{-\infty}dx_{N-1}...dx_1
~P[x_N,..,x_1|x_0] \nonumber \\
&\times&\exp\left[-\frac{\left({\cal I}-\varepsilon\sum_{n=1}^N
\alpha(x_n,x_{n-1};\varepsilon)\right)^2}{2\varepsilon^2
\sum_{n=1}^N\beta(x_n,x_{n-1};\varepsilon)}\right]
\left(2\pi\varepsilon^2 \sum_{n=1}^N\beta(x_n,x_{n-1};\varepsilon)
\right)^{-1/2}.
\end{eqnarray}
where $x_N\equiv x_u$, $x_t\equiv x_0$ and the discrete path {\sc pdf} 
is given by
\begin{equation}
P[x_N,...,x_1|x_0]=
\left(\frac{1}{2\pi\sigma^2\varepsilon}\right)^{\frac{N}{2}}
\exp\left[-\sum_{n=1}^N\frac{(x_n-x_{n-1}-\mu\varepsilon)^2}
{2\sigma^2\varepsilon}\right].
\end{equation}
Equation (3.11) describes the probability density of realizing a particular 
discrete path of the risk-neutral stochastic process (2.2).
In the limit that $\beta$ tends to zero, the joint {\sc pdf} (3.10) 
becomes a delta function in ${\cal I}$ and we can show, using (2.8), 
that the option price becomes
\begin{equation}
C_F(S_t,t)\simeq e^{-rT}\int_{-\infty}^{\infty}dx_{N}...dx_1
P[x_N,...,x_1|x_0]~F\Bigl(e^{x_N},\varepsilon\sum_{n=1}^N
\alpha(x_n,x_{n-1};\varepsilon)\Bigr)
+o(\varepsilon^2\sigma^2 T).
\end{equation}
The order of the correction term follows from (3.10) and a saddle point 
expansion of the Gaussian integral over ${\cal I}$ in (2.8). 

Equation (3.12) is the major result of this paper. The multi-dimensional 
integral can be evaluated by deterministic methods or 
more generally by standard Monte Carlo methods. In this case we use the 
observation that (3.12) is equivalent 
to an expectation of the option payoff function $F$, with respect to the 
discretely sampled risk-neutral diffusion process defined by (3.11), or 
equivalently by (2.2). In (3.12), the discretization
time-scale $\varepsilon$ is independent of any discrete option sampling 
time-scale.
If we choose $\varepsilon$ to match the interval between discrete 
option sampling, we find that (3.7) will reduce to a primitive short-time 
propagator and (3.12) will become
\begin{equation}
C_F(S_t,t)= e^{-rT}\int_{-\infty}^{\infty}dx_{N}...dx_1
P[x_N,...,x_1|x_0]
F\left(e^{x_N},\sum_{n=0}^N w_nf(x_n)\right).
\end{equation} 
This is equivalent to a direct discretization of (2.7), consistent with 
a discretely sampled path dependent random variable described by 
(2.5) and (2.6). In this case no analytical partial averaging has been performed 
and the Monte Carlo evaluation of (3.13) is completely standard. 
The great advantage of the partial averaging method is that 
in (3.12), the discrete time interval $\varepsilon$ can be chosen to be 
much larger than the option sampling time-scale.  
The partial averaging is performed in
(3.7) where we must average over Gaussian fluctuations about the 
straight line path $\bar{x}_{\tau}$ connecting $x_n$ and $x_{n-1}$.
This corresponds to averaging over the high frequency fluctuations of the 
risk-neutral diffusion process. In practice, as will be seen in the next 
section, the partial averaging integral is easily performed analytically. 
It is simply the Gaussian transform of the function $f$, which defines the 
path dependent random variable in question via (2.5). 
Of most practical interest will be discretely sampled path dependent options,
where $w(s)$ is given by (2.6). In this case the subsequent integral 
over $\tau$ in (3.7) reduces to a discrete sum which presents no problems.

For the special case $N=1$, for which $\varepsilon=T$, $x_1\equiv x_u$
and $x_0\equiv x_t$, we find that (3.12) becomes
\begin{equation}
C_F(S_t,t)\simeq \frac{e^{-rT}}{\sqrt{2\pi\sigma^2 T}}
\int^{\infty}_{-\infty}dx_u\,
\exp\left[-\frac{(x_u-x_t-\mu T)^2}{2\sigma^2 T}\right]
F\Bigl(e^{x_u},T\alpha(x_u,x_t;T)\Bigr)+o(\sigma^2 T^3).
\end{equation}
This describes rather generally an approximate path dependent 
option price. 
It can be simply evaluated as a one dimensional numerical integral. 
As a measure of the accuracy of (3.14), we ask at what time to maturity 
$T$, does the error term in (3.14) become $1\%$ of the true option price. 
We assume a typical market volatility of $\sigma=0.25$ and that the 
unknown coefficient of the correction term is equal to the true 
option price. We find that $T\simeq 0.5$ 
gives a $1\%$ error, while $T\simeq 0.25$ gives an error of approximately 
$0.1\%$. These estimates begin to illustrate the power of the 
method presented here.

\section{Examples}
The previous formulation is rather general and can be
applied to a range of path dependent options. In this section we will
show how two important and qualitatively different classes of path 
dependent options fit into this framework.

\subsection{Average rate options}
The payoff of the geometric Asian option is some function of the
path dependent random variable given by
\begin{equation}
{\cal I}=\int^u_t ds\,w(s)x_s~, 
\end{equation}
where $x_s$ is related to the risk-neutral asset price by (2.3). 
From (2.5) we can identify $f$ with $x_s$. 
After performing the partial averaging (3.7), we find
\begin{equation}
\alpha(x_n,x_{n-1};\varepsilon)=
\int_0^1d\tau\,w(\tau)\bar{x}_{\tau}.
\end{equation}
For the continuous sampling ($w(\tau)=1$) we find
\begin{equation}
\alpha(x_n,x_{n-1};\varepsilon)=(x_n+x_{n-1})/2.
\end{equation}
For this simple example $\alpha$ is just the primitive short-time propagator.

The payoff of the arithmetic Asian option will be some function of the
path dependent random variable 
\begin{equation}
{\cal I}=\int^u_t ds\,w(s)e^{x_s}.
\end{equation} 
From (2.5) we identify $f$ with $e^{x_s}$. 
After performing the partial averaging (3.7) we find
\begin{equation}
\alpha(x_n,x_{n-1};\varepsilon)=
\int_0^1d\tau\,w(\tau)e^{\bar{x}_{\tau}+\nu_{\tau}^2/2}.
\end{equation}
We can expand (4.5) to order $\sigma^2\varepsilon$ without a significant 
loss of accuracy. This is consistent with the order of the truncation 
of the cumulant expansion (3.5). We then find
\begin{equation}
\alpha(x_n,x_{n-1};\varepsilon) \simeq
\int_0^1d\tau\,w(\tau)e^{\bar{x}_{\tau}}\left(1+\sigma^2\varepsilon(1-\tau)\tau/2
+o(\sigma^4\varepsilon^2)\right).
\end{equation}
The final result will depend on whether we use discrete or continuous sampling.
For continuous sampling we have $w(\tau)=1$ and (4.6) becomes
\begin{equation}
\alpha(x_n,x_{n-1};\varepsilon)\simeq 
e^{x_{n-1}}\left[\frac{1}{a}\Bigl(e^a -1\Bigr)+
\frac{\sigma^2\varepsilon}{2a^3}
\Bigl(e^a(a-2)+a+2\Bigr)+o(\sigma^4\varepsilon^2)\right],
\end{equation}
where $a=x_n-x_{n-1}$.
For discrete sampling we can perform the necessary 
summations analytically so the method is equally effective. 
It is instructive to compare (4.7) with the $\alpha$ that generates
the primitive short-time propagator. This is obtained by approximating 
(3.7) by 
\begin{equation}
\alpha(x_n,x_{n-1};\varepsilon)\simeq 
\frac{1}{2}\Bigl(f(x_n)+f(x_{n-1})\Bigr).
\end{equation}
One can see that (4.7) will only reduce to (4.8) in the limit 
$a\rightarrow 0$ (when $f=e^{x_s}$).

We can now use (3.12) to perform a Monte Carlo evaluation 
of the continuously sampled Asian option. Using (4.7) allows us to 
obtain accurate results by simulating 
only relatively low dimensional random paths of (2.2). 
This will avoid the problems associated with the 
Monte Carlo evaluation of these options \cite{mad}.

\subsection{Occupation time derivatives}
A large and important class of path dependent options are those where 
the payoff depends on the time the asset price spends within a given region. 
This class of options have been referred to as 
occupation time derivatives and they have been discussed
in detail by Linetsky \cite{linetsky2,linetsky3} 
and Hugonnier \cite{hug}. The valuation of debt and contingent claims 
with default risk can also be placed in this class \cite{lon,ric}.

As a simple example, consider an option with a payoff that depends on the
time spent below a possibly time dependent barrier $B_s$. This time is 
a path dependent random variable which can be expressed as 
(and similarly for the above case)
\begin{equation}
{\cal I}=\int_t^u ds\,w(s)H(B_s-e^{x_s}),
\end{equation}
where $H$ is a simple step function. We can therefore identify
$H$ with the function $f$ in (2.5). After performing the partial 
averaging (3.7) we find
\begin{equation}
\alpha(x_n,x_{n-1};\varepsilon)=
\int_0^1d\tau\,w(\tau) N\Bigl((\ln B_s -\bar{x}_{\tau})/\nu_{\tau}\Bigr),
\end{equation}
where $N$ is the cumulative normal distribution function. 
This function is difficult to integrate analytically. However the 
real case of practical interest is discrete sampling. In this case 
the integration reduces to a discrete summation and presents no 
problems. Barrier options are the most basic and well known example 
of options that fall under this framework. An interesting generalization 
is the step option which has a finite knock-out rate. This is motivated by 
risk-management arguments and a variety of possible payoff functions 
have been discussed \cite{linetsky2,linetsky3}. All variations are 
included in this framework since the payoff function is an arbitrary 
function of the path dependent random variable.

Consider the example of derivatives depending on the occupation time 
between two barriers $B_1$ and $B_2$.\footnote{the occupation time 
outside these barriers can be trivially constructed from this time}
Included in this class are 
double barrier options \cite{ger} and the range products such as 
range notes and corridor options \cite{pec,tur}.
This time is a path dependent random variable which can be expressed as
\begin{equation}
{\cal I}=\int_t^u ds\,w(s)\Bigl[H(B_2-e^{x_s})-H(B_1-e^{x_s})\Bigr].
\end{equation}
From (2.5) we can identify the function $f$ with the difference of two step
functions. After performing the partial averaging (3.7) we find
\begin{equation}
\alpha(x_n,x_{n-1};\varepsilon)=
\int_0^1d\tau\,w(\tau)
\Bigl[ N\Bigl((\ln B_2 -\bar{x}_{\tau})/\nu_{\tau}\Bigr)
-N\Bigl((\ln B_1 -\bar{x}_{\tau})/\nu_{\tau}\Bigr)\Bigr].
\end{equation}
Our framework is easily generalized to include derivatives dependent 
on several occupation times.

\section{Discussion and Conclusion}
The aim of this paper was to present a new approach to evaluating 
the price of path dependent options. We considered options with 
the general payoff function (2.4), contingent on a path 
dependent random variable 
expressible in the form (2.5). The key results of this paper were 
the general evaluation formula (3.12) and the short time to expiry 
approximation (3.14). They give the option price after 
performing analytically a partial averaging, defined by equations 
(3.7-9), over the underlying risk-neutral diffusion process (2.2). 
Since the method is analytically based, it also gives the order of the 
error made by choosing a finite time discretization.
Specific examples were presented in section 4.

In (3.12), the partial averaging allows one to choose the discrete 
time interval $\varepsilon$ to be much larger than the option 
sampling time-scale. 
We could, for example, imagine choosing $\varepsilon$ to be one month 
for an option with daily sampling. 
Clearly, the partial averaging method can greatly reduce the dimension 
of the integral in (3.12). This integral can be evaluated most generally by
standard Monte Carlo methods. In this case the partial averaging 
will greatly reduce the dimension of the random paths to be simulated.
In effect, it allows random simulations to be replaced with 
deterministic calculations. Standard methods to increase the efficiency of 
the Monte Carlo evaluation can still be used. 
These include variance reduction techniques \cite{cle}, the simulation of 
sample paths using the Brownian bridge process \cite{jung} or the use 
of quasi Monte Carlo sampling \cite{cle,pap}. Interestingly, 
quasi Monte Carlo sampling is known to be more advantageous 
for low dimensional numerical integrals. The partial averaging method 
can therefore increase the relative gains made by the 
implementation of quasi Monte Carlo methods.

The framework presented here can be extended to multi-factor path dependent 
options \cite{linetsky1,rosa-clot1} and to path dependent options dependent 
on several path dependent random variables. Recent work has shown 
that a path-independent option price, in a stochastic volatility environment,
can be approximated by pricing a more complex path dependent option in the 
usual Black-Scholes framework. The volatility 
risk is included by an extra continuous path-dependent payout function 
which has the same form as (2.5) \cite{fouque1,fouque2}. 
This work suggests that the computational framework presented here might 
also be used to price path dependent options consistent with the 
market implied volatility. Further work is required in this direction.

Path integral methods have 
long been developed and used as a computational tool in theoretical and 
chemical physics. Hopefully the work presented here will stimulate more 
interest in the application of these methods to problems in 
computational finance.

\appendix
\section{Path Integral Representation}
Consider a stochastic process $x_s$, which obeys the stochastic differential 
equation \footnote{
Note that any one-dimensional risk-neutral diffusion process 
can be cast into this form by a change of variable}
\begin{equation}
dx_{t}=-g^{\prime }(x_{t})dt+\sigma dW_{t}.
\end{equation}
In our notation $x_s\equiv x(s)$.
We will rewrite this equation as
\begin{equation}
\dot{x}_{t}=-g^{\prime }(x_{t})+\sigma \zeta _{t}
\end{equation}
where $\zeta _{t}=\frac{dW_{t}}{dt}$ and a dot denotes
the derivative with respect to time. This notation is more
suited for the path integral formulation. In (A.1) we assume that 
$\sigma$ in independent of $x_t$ (additive noise). The path integral 
formulation of the more general case (multiplicative noise) can 
be found in \cite{arnold}.

Consider a general continuous Gaussian noise process $\chi(s)$ (note
that small $s$ denotes the time history variable between time $t$ and
$u$ and should not be confused with the price $S_{t}$). We discretize
this process by defining a discrete time $s_{n}=n\varepsilon +t$,
where $n=0,1,...,N$ and $\varepsilon =T/N$ with $T=u-t$. Note that
$\varepsilon $ is equivalent to $ds$ when $N\rightarrow
\infty $. The now discrete Gaussian process is fully defined by the
normalized probability density functional
\begin{equation}
{\cal P}\left[ \chi_{1},...,\chi_{N}\right] =\left( 2\pi \right) ^{-n/2}
\left( \det R\right) ^{-1/2}\exp
\left[ -\frac{1}{2}\sum_{n,m=1}^{N}\chi_{n}\,R_{nm}^{-1}\,\chi_{m}
\right],
\end{equation}
where $\chi_{n}=\chi(s_{n})$, $E[\chi_{n}\chi_{m}]=R_{nm}$
and $R_{nm}^{-1}$ denotes the inverse matrix. Equation (A.3) fully defines
the probability density of the whole history of the discrete Gaussian process.
This normalization condition implies
\begin{equation}
\int^{\infty}_{-\infty}
{\cal P}\left[ \chi_{1},...,\chi_{N}\right]\,d\chi_1...d\chi_N=1.
\end{equation}
Since $dW_{t}$ has a variance $dt$, we know $\zeta _{t}$ will have a variance $dt^{-1}$.
We then find that for the discrete time white
noise process $\zeta _{n}$, we have $E[\zeta _{n}\zeta
_{m}] =\varepsilon ^{-1}\delta _{nm}$ where $\delta _{nm}$ is
the unit diagonal matrix. Using (A.3) we then find
\begin{equation}
{\cal P}\left[ \zeta _{1},...,\zeta _{N}\right] =\left( \frac{\varepsilon }{2\pi }%
\right) ^{N/2}\exp \left[ -\frac{1}{2}\sum_{n=1}^{N}\varepsilon \,\zeta
_{n}^{2}\right].
\end{equation}
In the continuous limit this becomes
\begin{equation}
{\cal P}[\zeta_s]=\left( \frac{\varepsilon }{2\pi }\right) ^{N/2}
\exp \left[ -\frac{1}{2}\int_{t}^{u}ds\,\,\zeta ^{2}_s\,\right],
\end{equation}
where the continuous time expressions are written with the
understanding that $\varepsilon \rightarrow 0$ and $N\rightarrow \infty $
such that $N\varepsilon =T$.

We wish to use the probability density functional (A.6) to
find a probability density functional for the stochastic process
$x_s$. To do this we need to first discretize in time (A.2). 
In discrete time (A.2) becomes
\begin{equation}
\frac{x_{n-}x_{n-1}}{\varepsilon }=-g^{\prime }(\tilde{x}_{n})+\sigma \zeta
_{n},
\end{equation}
where
\begin{equation}
\tilde{x}_{n}=\phi x_{n}+(1-\phi )x_{n-1}
\end{equation}
and $\phi $ is a discretization parameter between 0 and 1.
From (A.7) we see
that a path $\left\{ \zeta _{1},...,\zeta _{N}\right\} $
maps to a unique path $\left\{ x_{1},...,x_{N}\right\} $
\emph{as long as }$x_{0}$\emph{\ is given}.
We will therefore write, by virtue of (A.4)
\begin{equation}
\int_{-\infty}^{\infty}{\cal P}[x_1,...,x_N|x_0]\,{\cal J}\,dx_1...dx_N=1,
\end{equation}
where ${\cal J}$, defined by
\begin{equation}
{\cal J}={\rm det}\left|\frac{\partial \zeta _{n}}{\partial x_{m}}\right|
,\;\;\;m,n=1,...,N
\end{equation}
is the Jacobian of the change in coordinates.
In the continuous limit we can substitute (A.2) into (A.6) to obtain
\begin{equation}
{\cal P}\left[ x_s\right] \,=\left( \frac{\varepsilon }{2\pi }\right)
^{N/2}\exp \left[ -\frac{1}{2\sigma ^{2}}\int_{t}^{u}ds\,\,
\Bigl( \dot{x}_s+g^{\prime }(x_s)\Bigr)^{2}\,\right].
\end{equation}
We show in appendix B that the Jacobian (A.10) becomes
in the continuous limit
\begin{equation}
{\cal J}=\left( \sigma \varepsilon \right) ^{-N}\exp \left[ \phi
\int_{t}^{u}ds\,\,g^{\prime \prime }\left( x_s\right) \,\right].
\end{equation}
The only sensible choice is $\phi =1/2$ \cite{arnold}. 
Combining the Jacobian and
(A.11) we obtain a new probability density functional
\begin{equation}
{\cal P}_x\left[ x_s\right] \,=\exp \left[ -\frac{1}{2\sigma ^{2}}
\int_{t}^{u}ds\,\,\Bigl( \dot{x}_s+g^{\prime }(x_s)\Bigr)^{2}\,
+\frac{1}{2}\int_{t}^{u}ds\,\,g^{\prime\prime }(x_s)\right],
\end{equation}
which is normalized with respect to the functional measure
${\cal D}_s$ which is the continuous limit of
\begin{equation}
{\cal D}x_s=\,\left( 2\pi \varepsilon \sigma ^{2}\right)
^{-N/2}dx_{1}...dx_{N-1}.
\end{equation}
This means the conditional probability density function ({\sc pdf}) 
of (A.2) is given by the path integral
\begin{equation}
P(x_{u},x_{t})=\int_{x_{t}}^{x_{u}}{\cal D}x_s\,{\cal P}_x[x_s].
\end{equation}
The expectation of a functional ${\cal F}[x_s]$, conditional on the
initial value $x_t$, can now be expressed in the path integral form
\begin{equation}
E_{x_t}\Bigl[{\cal F}[x_s]\Bigr]=\int^{\infty}_{-\infty}dx_u
\int_{x_{t}}^{x_{u}}{\cal D}x_s\,{\cal P}_x[x_s]\,{\cal F}[x_s].
\end{equation}
Using (2.7), we can now write the option price in the path integral form
\begin{equation}
C_F(S_t,t)=e^{-rT}\int^{\infty}_{-\infty}dx_u
\int_{x_{t}}^{x_{u}}{\cal D}x_s\,{\cal P}_x[x_s]\,F(e^{x_u},{\cal I}).
\end{equation}
We will not deal directly with the above path integral representation 
of the option price. From (2.8) we see that we can extract the 
payoff function out of the path integral if we instead focus on the 
path integral representation of the joint {\sc pdf} $P(x_u,{\cal I}|x_t)$. 
In the next subsection we will see how to calculate this function.

\subsection{Calculating the joint {\sc pdf}}
The joint {\sc pdf} introduced in (2.8) can be simply expressed 
as the path integral
\begin{equation}
P(x_u,{\cal I}|x_t)=\int_{x_t}^{x_u}{\cal D}\hat{x}_s\,{\cal P}_x[\hat{x}_s]
\delta({\cal I}-\hat{{\cal I}}).
\end{equation}
Using the Fourier representation of the delta function and (2.5), we find that
(A.18) becomes
\begin{equation}
P(x_u,{\cal I}|x_t)=\frac{1}{2\pi}\int_{-\infty}^{\infty}dk
e^{-ik{\cal I}}
\int_{x_t}^{x_u}{\cal D}\hat{x}_s\,{\cal P}_x[\hat{x}_s]
\exp\left[ik\int_t^u ds\,w(s)f(\hat{x}_s,s)\right].
\end{equation}
Substituting (A.13) into (A.19) and using
\begin{equation}
\int_{t}^{u}ds\,\dot{x}_s\,g^{\prime }(x_s)=g(x_{u})-g(x_{t}),
\end{equation}
we find that (A.19) becomes
\begin{equation}
P(x_u,{\cal I}|x_t)=\frac{1}{2\pi}\exp\left[\frac{g(x_t)-g(x_u)}{\sigma^2}
\right]\int_{-\infty}^{\infty}dk e^{-ik{\cal I}}K(x_u,x_t;T),
\end{equation}
where $K$, which we will refer to as the propagator, is defined by
\begin{equation}
K(x_{u},x_{t};T)=\int_{x_{t}}^{x_{u}}{\cal D}x_s\exp
\left[ -\frac{1}{2\sigma ^{2}}\int_{t}^{u}ds\,\,\left[ \dot{x}^{2}_s
+V(x_s,s)\right] \right]
\end{equation}
and what we will call the potential function $V$ is
\begin{equation}
V(x_s,s)=g^{\prime 2}(x_s)-\sigma ^{2}\,\,g^{\prime \prime}(x_s)
-2ik\sigma ^{2}\,w(s)\,f(x_s,s).
\end{equation}
Of special interest is the geometric Brownian motion 
model defined by (2.2). 
Comparing (2.2) with (A.1), we can identify $g^{\prime}(x_t)$ with 
$\mu$ and we find the joint {\sc pdf} (A.21) becomes (2.9) with the 
potential (2.11). Because the drift term is constant in this case, it can 
be extracted out of the path integral and we are left with an imaginary 
potential.

\section{Calculating the Jacobian}
In this appendix we will calculate the Jacobian (A.10).
In discrete time the stochastic differential equation ({\sc sde}) 
(A.2) becomes
\begin{equation}
\frac{x_{n-}x_{n-1}}{\varepsilon }=-g^{\prime }(\tilde{x}_{n})+
\sigma(\tilde{x}_n)\zeta_{n},
\end{equation}
where it has been generalized to include a non-constant diffusion coefficient.
From this discrete {\sc sde} we see that
\begin{equation}
\frac{\partial \zeta_n}{\partial x_m}=0,\,\,\,{\rm for}\,\,\,m>n.
\end{equation}
Therefore the Jacobian matrix is triangular and from (A.10) we obtain
\begin{equation}
{\cal J}=\prod_{n=1}^N\frac{\partial \zeta_n}{\partial x_n}.
\end{equation}
From (B.1) we find
\begin{equation}
\frac{1}{\varepsilon}\frac{\partial x_n}{\partial \zeta_n}=-\phi
g^{\prime \prime}(\tilde{x}_n)\frac{\partial x_n}{\partial \zeta_n}
+ \phi\,\zeta_n\,\sigma^{\prime}(\tilde{x}_n)\frac{\partial x_n}{\partial \zeta_n}
+\sigma(\tilde{x}_n),
\end{equation}
where we have used
\begin{equation}
\frac{\partial \tilde{x}_n}{\partial \zeta_n}=
\frac{\partial \tilde{x}_n}{\partial x_n}
\frac{\partial x_n}{\partial \zeta_n}=
\phi\frac{\partial x_n}{\partial \zeta_n}
\end{equation}
obtained from (A.8).
Multiplying (B.4) by $\varepsilon\frac{\partial \zeta_n}{\partial x_n}$ we obtain
\begin{equation}
\frac{\partial \zeta_n}{\partial x_n}=\frac{1+\varepsilon\,\phi
g^{\prime \prime}(\tilde{x}_n)-\varepsilon\,\phi\,\zeta_n\sigma^{\prime}
(\tilde{x}_n)}{\varepsilon\sigma(\tilde{x}_n)}.
\end{equation}
For small $\varepsilon$ (B.6) becomes
\begin{eqnarray}
\frac{\partial x_n}{\partial \zeta_n}&\simeq& \varepsilon\sigma(\tilde{x}_n)
\Bigl(1-\varepsilon\,\phi\,g^{\prime \prime}(\tilde{x}_n)+
\varepsilon\,\phi\,\zeta_n\sigma^{\prime}(\tilde{x}_n)\Bigr) \nonumber \\
&\simeq& \varepsilon\sigma(\tilde{x}_n)\exp\Bigl[\varepsilon\,
\phi\Bigl(\sigma^{\prime}(\tilde{x}_n)\zeta_n-g^{\prime \prime}(\tilde{x}_n)
\Bigr)\Bigr].
\end{eqnarray}
Using this result we find
\begin{equation}
\prod_{n=1}^N\frac{\partial x_n}{\partial \zeta_n}\simeq
\varepsilon^N\left(\prod_{n=1}^N\sigma(\tilde{x}_n)\right)
\exp\left[\varepsilon\,\phi\sum_{n=1}^N
\Bigl(\sigma^{\prime}(\tilde{x}_n)\zeta_n-g^{\prime \prime}(\tilde{x}_n)
\Bigr)\right].
\end{equation}
In the continuous limit $\varepsilon\rightarrow 0$, we therefore find
\begin{equation}
{\cal J}^{-1}=\varepsilon^N\left(\prod_{n=1}^N\sigma(\tilde{x}_n)\right)
\exp\left[\phi\int_{t}^u ds\,
\Bigl(\sigma^{\prime}(x_s)\zeta_s-g^{\prime \prime}(x_s)
\Bigr)\right].
\end{equation}
From the continuous {\sc sde} (A.2) we have
\begin{equation}
\zeta_s=\Bigl(\dot{x}_s+g^{\prime}(x_s)\Bigr)/\sigma(x_s).
\end{equation}
Substituting this into (B.9) we find
\begin{equation}
{\cal J}=\varepsilon^{-N}\left(\prod_{n=1}^N\sigma(\tilde{x}_n)\right)^{-1}
\exp\left[-\phi\int_{t}^u ds\,
\Bigl((\dot{x}_s+g^{\prime}(x_s))\sigma^{\prime}(x_s)/\sigma(x_s)-
g^{\prime \prime}(x_s)\Bigr)\right].
\end{equation}
This agrees with equation 2.4.37 in Stratonovich \cite{strat} who
derives the Jacobian for a general multi-factor system of {\sc sde}'s.
For a constant diffusion coefficient (B.11) reduces to (A.12).

\section{Improved Short-time Propagator}
The goal of this appendix is to obtain an expansion of
$K(x_u,x_t;T)$ in powers of $T$. The derivation here is inspired by the
cumulant method used in connection with the quantum statistical density
matrix \cite{coalson}.

We must first write the propagator in the Fourier path integral 
representation ({\sc fpir}).
We decompose the paths as
\begin{equation}
x_\tau=\bar{x}_{\tau}+\left(\frac{4\sigma^2 T}{\pi}\right)^{1/2}
\sum_{n=1}^{\infty} \frac{z_n\sin(n\pi\tau)}{n}
\end{equation}
where
\begin{equation}
\bar{x}_{\tau}=\tau(x_u-x_t)+x_t,\,\,\,\tau=(s-t)/T.
\end{equation}
The term $\bar{x}_{\tau}$ in (C.1) is the straight line path
connecting $x_t$ and $x_u$. The remaining terms are the harmonic
perturbations about the straight line path. With this we can
now write
\begin{equation}
\frac{1}{2\sigma^2}\int_{t}^{u}ds\,\left[ \dot{x}^{2}_s
+V(x_s,s)\right]=\frac{(x_u-x_t)^2}{2\sigma^2 T}+
\pi\sum_{n=1}^{\infty}z_n^2 +
\frac{T}{2\sigma^2}\int_0^1d\tau\,V(x_{\tau},\tau).
\end{equation}
Summing over all paths
between $x_t$ and $x_u$ is equivalent to integrating over all possible
values of the Fourier coefficients $\{z_n\}$. This means we can write
\begin{equation}
\int_{x_t}^{x_u}{\cal D}x_s=({\rm constant})\times
\int_{-\infty}^{\infty}dz_1...dz_{\,\infty},
\end{equation}
where the constant is some Jacobian factor resulting from the
change of integration variables. Substituting (C.4) and (C.3)
into (2.10), we obtain the {\sc fpir} of the propagator
\begin{equation}
K(x_u,x_t;T)=K_f(x_u,x_t;T)\int_{-\infty}^{\infty}dz_1...dz_{\,\infty}
\exp\left[-\pi\sum_{n=1}^{\infty}z_n^2 -
\frac{T}{2\sigma^2}\int_0^1d\tau\,V(x_{\tau},\tau)\right]
\end{equation}
where $K_f(x_u,x_t;T)$ is given by
\begin{equation}
K_f(x_u,x_t;T)=\left(\frac{1}{2\pi\sigma^2 T}\right)^{1/2}
\exp\left(-\frac{(x_u-x_t)^2}{2\sigma^2 T}\right).
\end{equation}
The constant Jacobian factor in (C.4) must equal the prefactor of (C.6)
to obtain the correct propagator when $V=0$ (clearly the
Jacobian is independent of the potential).
We can rewrite (C.5) as
\begin{equation}
K(x_u,x_t;T)=K_f(x_u,x_t;T)\,E\left[
\exp\left(-\frac{T}{2\sigma^2}\int_0^1d\tau\,V(x_{\tau},\tau)\right)
\right]_{\{z_n\}},
\end{equation}
where the expectation is with respect to the set of independent
Gaussian random variables ${\{z_n\}}$ with zero mean and
variance $1/2\pi$. The {\sc fpir} can also be set up using a reference potential that is
quadratic \cite{miller}.

We can use the {\sc fpir} (C.7) to derive an expansion in time for the
short time propagator. We have using the cumulant expansion
\begin{equation}
K(x_u,x_t;T)=K_f(x_u,x_t;T)
\exp\left(\sum_{m=1}^{\infty}\frac{1}{m!}\left(-\frac{T}{2\sigma^2}
\right)^m C_m(x_u,x_t;T)\right),
\end{equation}
where
\begin{equation}
C_1=M_1,\,\,C_2=M_2-M_1^2,\,\, C_3=M_3-3M_1M_2+2M_1^3,...
\end{equation}
and
\begin{equation}
M_m=E\left[\left(\int_0^1d\tau\,V(x_{\tau},\tau)\right)^m
\right]_{\{z_n\}}.
\end{equation}
The propagator (C.8) has the general structure defined by (3.2) and (3.5).
Consider the first cumulant which is
\begin{equation}
C_1(x_u,x_t;T)=\int_0^1d\tau\,E\left[V(x_{\tau},\tau)\right]_{\{z_n\}}.
\end{equation}
We know from (C.1) that $x_{\tau}$ is a sum of Gaussian random variables
which itself must be Gaussian distributed.
So when calculating the expectation (C.11), we can replace
$x_{\tau}$ by
\begin{equation}
x_{\tau}=\bar{x}_{\tau}+p_{\tau},
\end{equation}
where $p_{\tau}$ is a single Gaussian random variable with variance
\begin{equation}
\nu_\tau^2=\frac{2\sigma^2T}{\pi^2}\sum_{n=1}^\infty
\frac{\sin^2(n\pi\tau)}{n^2}=\sigma^2 T(1-\tau)\tau.
\end{equation}
Using this result we find that
\begin{equation}
E\left[V(x_{\tau},\tau)\right]_{\{z_n\}}=
\int^{\infty}_{-\infty}dp_{\tau}\,P(p_{\tau})
V(\bar{x}_\tau+p_{\tau},\tau)
\end{equation}
where
\begin{equation}
P(p_{\tau})=\frac{1}{\sqrt{2\pi\nu_\tau^2}}
\exp(-p_{\tau}^2/2\nu_\tau^2).
\end{equation}
The first cumulant then becomes
\begin{equation}
C_1(x_u,x_{t};T)=
\int_0^1d\tau\,\int^{\infty}_{-\infty}dp_{\tau}\,P(p_{\tau})
V(\bar{x}_\tau+p_{\tau},\tau).
\end{equation}
Expanding the potential around $p_{\tau}=0$ and performing the
Gaussian integrals we obtain
\begin{equation}
E\left[V(x_{\tau},\tau)\right]_{\{z_n\}}\simeq V(\bar{x}_\tau,\tau)
+\frac{\nu_{\tau}^2}{2}V^{\prime\prime}(\bar{x}_\tau,\tau)
+\frac{\nu_{\tau}^4}{8}V^{\prime\prime\prime\prime}(\bar{x}_\tau,\tau)
+o(T^3),
\end{equation}
which from (C.13) is an expansion in time $T$.

Consider now the second cumulant
\begin{equation}
C_2(x_u,x_t;T)=\int^1_0d\tau d\tau'\Biggl(E\left[V(x_{\tau},\tau)
V(x_{\tau'},\tau')\right]_{\{z_n\}}-
E\left[V(x_{\tau},\tau)\right]_{\{z_n\}}
E\left[V(x_{\tau'},\tau')\right]_{\{z_n\}}\Biggr).
\end{equation}
In (C.18) we can replace $x_{\tau}$ and $x_{\tau'}$ with
\begin{equation}
x_{\tau}=\bar{x}_{\tau}+p_{\tau},\;\;
x_{\tau'}=\bar{x}_{\tau'}+p_{\tau'}
\end{equation}
where $p_{\tau}$ and $p_{\tau'}$ are two correlated Gaussian
random variables with variances
\begin{equation}
\nu_\tau^2=\sigma^2 T(1-\tau)\tau,\;\;\;
\nu_{\tau'}^2=\sigma^2 T(1-\tau')\tau'.
\end{equation}
Using (C.1) we find the covariance is
\begin{equation}
c(\tau,\tau')=E[p_{\tau}p_{\tau'}]=\frac{4\sigma^2 T}{\pi}
\sum_{n,n'=1}^{\infty}E[z_n z_{n'}]_{\{z_n\}}
\frac{\sin(n\pi\tau)\sin(n'\pi\tau')}{nn'}.
\end{equation}
Using
\begin{equation}
E[z_n z_{n'}]_{\{z_n\}}=\frac{1}{2\pi}\delta_{nn'}
\end{equation}
and the identity
\begin{equation}
\frac{2}{\pi^2}\sum_{n=1}^{\infty}\frac{\sin(n\pi\tau)\sin(n\pi\tau')}{n^2}
=\tau_l(1-\tau_g),
\end{equation}
we find
\begin{equation}
c(\tau,\tau')=\sigma^2 T\,\tau_{l}(1-\tau_g)
\end{equation}
where $\tau_l$ is the lesser of $\tau$ and $\tau'$ and
$\tau_g$ is the greater of $\tau$ and $\tau'$.
Using (A.3), we can write the probability distribution of $p_{\tau}$
and $p_{\tau'}$ as
\begin{equation}
P(p_{\tau},p_{\tau'})=\frac{1}{2\pi}
\frac{1}{\sqrt{\nu_{\tau}^2\nu_{\tau'}^2-c^2}}
\exp\left[-\frac{(\nu_{\tau'}^2 p_{\tau}^2-2c \,p_{\tau}p_{\tau'}
+\nu_{\tau}^2p_{\tau'}^2)}{2(\nu_{\tau}^2\nu_{\tau'}^2-c^2)}\right].
\end{equation}
We therefore find that
\begin{equation}
E\left[V(x_{\tau},\tau)V(x_{\tau'},\tau')\right]_{\{z_n\}}=
\int_{-\infty}^{\infty}dp_{\tau}dp_{\tau'}P(p_{\tau},p_{\tau'})
V(\bar{x}_{\tau}+p_{\tau},\tau)V(\bar{x}_{\tau'}+p_{\tau'},\tau').
\end{equation}
Substituting this result and (C.14) into (C.18), we find that the second
cumulant is
\begin{eqnarray}
C_2(x_u,x_{t};T)&=&
\int_0^1d\tau d\tau'\left(
\int_{-\infty}^{\infty}dp_{\tau}dp_{\tau'}
\,P(p_{\tau},p_{\tau'})
V(\bar{x}_{\tau}+p_{\tau},\tau)V(\bar{x}_{\tau'}+p_{\tau'},\tau')
\right. \nonumber \\
&-& \left. \int_{-\infty}^{\infty}dp_{\tau}dp_{\tau'}
\,P(p_{\tau})P(p_{\tau'})
V(\bar{x}_{\tau}+p_{\tau},\tau)
V(\bar{x}_{\tau'}+p_{\tau'},\tau')\right).
\end{eqnarray}
Expanding the second cumulant around $p=0$ we find
that the first order remaining term is
\begin{equation}
C_2(x_u,x_t;T)\simeq\sigma^2 T\int^1_0d\tau d\tau'
\,V^{\prime}(\bar{x}_{\tau},\tau)V^{\prime}(\bar{x}_{\tau'},\tau')
\,\tau_{l}(1-\tau_g)+o(\sigma^4 T^2 ).
\end{equation}
This follows since the integral we have to do is just the covariance
$c$.

It can be shown that the $m$th cumulant is of order $(\sigma^2 T)^{m-1}$.
Therefore from (C.8) we see that $C_2$ starts contributing
at order $T^3$, while $C_3$ starts contributing at order $T^5$.
This means that to get the expansion (C.8) correct to order $T^3$, we need
only expand $C_1$ to order $T^2$ and $C_2$ to order $T$ as has been done in
(C.17) and (C.28). This means that $C_1$ alone will give the correct propagator
to order $T^2$.
Substituting (C.28) and (C.17) into (C.8) we find
\begin{eqnarray}
K(x_u,x_t;T)&\simeq&K_f(x_u,x_t;T)
\exp\left[-\frac{T}{2\sigma^2}\int_0^1d\tau\,V(\bar{x}_\tau,\tau)
-\frac{T^2}{4}\int_0^1d\tau\,\tau(1-\tau)\,V^{\prime\prime}
(\bar{x}_\tau,\tau)\right. \nonumber \\
&-&\left.\frac{\sigma^2 T^3}{16}\int^1_0d\tau\,\tau^2(1-\tau)^2
V^{\prime\prime\prime\prime}(\bar{x}_\tau,\tau)\right. \nonumber \\
&+&\left.\frac{T^3}{8\sigma^2}\int^1_0d\tau d\tau'
\,V^{\prime}(\bar{x}_{\tau},\tau)V^{\prime}(\bar{x}_{\tau'},\tau')
\,\tau_{l}(1-\tau_g)+o(T^4)\right],
\end{eqnarray}
which is our desired expansion of the propagator to third order in time.

\vskip 1cm
\noindent{\bf Acknowledgements}:
I would like to thank the Australia Research Council for their
generous support of this research
through an Australian Postdoctoral Research Fellowship. I would also 
like to thank Science \& Finance and the School of Economics and 
Finance, University of Technology, Sydney, where part of this research 
was carried out.


\end{document}